\documentclass{amsart}
\usepackage{geometry}
\usepackage[pdftex]{hyperref}
\usepackage{graphics}
\usepackage{amscd}
\usepackage{xcolor,graphicx}

\newtheorem{theorem}{Theorem}[section]
\newtheorem{lemma}[theorem]{Lemma}
\newtheorem{pro}[theorem]{Proposition}
\newtheorem{cor}[theorem]{Corollary}

\theoremstyle{definition}

\newtheorem{example}[theorem]{Example}

\theoremstyle{remark}

\numberwithin{equation}{section}

\begin{document}
\pagestyle{plain}

\author[silas]{Silas L. Carvalho}
\address{Departamento de Matem\'atica, UFMG, Belo Horizonte, MG, 30161-970 Brazil.}
\email{silas@mat.ufmg.br}

\author[guidi]{Leonardo F. Guidi}
\address{Instituto de Matem\'atica e Estat\'istica, UFRGS, Porto Alegre, RS, 91509-900 Brazil.}
\email{guidi@mat.ufrgs.br}

\author[cfelipe]{Carlos F. Lardizabal}
\address{Instituto de Matem\'atica e Estat\'istica, UFRGS, Porto Alegre, RS, 91509-900 Brazil.}
\email{cfelipe@mat.ufrgs.br}

\def\ora{\overrightarrow}
\def\laa{\langle}
\def\raa{\rangle}
\def\qed{\begin{flushright} $\square$ \end{flushright}}
\def\qee{\begin{flushright} $\Diamond$ \end{flushright}}
\def\ov{\overline}

\date{\today}

\title{Site recurrence of open and unitary quantum walks on the line}


\begin{abstract}
We study the problem of site recurrence of discrete time nearest neighbor open quantum random walks (OQWs) on the integer line, proving basic properties and some of its relations with the corresponding problem for unitary (coined) quantum walks (UQWs). For both kinds of walks our discussion concerns two notions of recurrence, one given by a monitoring procedure \cite{werner,ls2015}, another in terms of P\'olya numbers \cite{stefanak}, and we study their similarities and differences. In particular, by considering UQWs and OQWs induced by the same pair of matrices, we discuss the fact that recurrence of these walks are related by an additive interference term in a simple way. Based on a previous result of positive recurrence we describe an open quantum version of Kac's lemma for the expected return time to a site.
\end{abstract}

\maketitle


\section{Introduction}

The model of {\bf coined (unitary) Quantum Random Walks} (UQW) has been widely studied in recent years and has found numerous applications in quantum information theory \cite{portugal,salvador}. The associated discrete time map acting on the integers can be written as
\begin{equation}U=S\cdot(C\otimes I),\end{equation}
where $U$ is a linear unitary operator acting on the Hilbert space given by the tensor product $\mathcal{H}_C\otimes\mathcal{H}_P$ of the so-called coin space, $\mathcal{H}_C$, and the state space, $\mathcal{H}_P$. The unitary map $C$ is defined as the coin, and the map $S$ is the shift operator
\begin{equation}S=|\uparrow\rangle\langle\uparrow|\otimes\sum_{i\in\mathbb{Z}}|i+1\rangle\langle i|+|\downarrow\rangle\langle\downarrow|\otimes\sum_{i\in\mathbb{Z}}|i-1\rangle\langle i|.\end{equation}
We consider $\mathcal{H}_C=\mathbb{C}^2$ and $\mathcal{H}_P=l_2(\mathbb{C})$ for 1-qubit walks on the integers. If we write
\begin{equation}\label{LR}
C=\begin{bmatrix} a & b \\ c & d \end{bmatrix}, \;\;\; R=\begin{bmatrix} a & b \\0 & 0 \end{bmatrix},\;\;\; L=\begin{bmatrix} 0 & 0 \\ c & d \end{bmatrix}\end{equation}
(so that $C$ is assumed unitary) and $|\psi\rangle\otimes|0\rangle$ is a state localized at $|0\rangle$, then a simple calculation shows that
\begin{equation}\label{canonicuqw}
U(|\psi\rangle\otimes|0\rangle)=R|\psi\rangle\otimes|1\rangle+L|\psi\rangle\otimes|-1\rangle,
\end{equation}
and the probabilities of finding the particle at site $|-1\rangle$ or $|1\rangle$ after one step are, respectively, $\Vert L|\psi\rangle\Vert^2$ and $\Vert R|\psi\rangle\Vert^2$. After two steps, $U^2(|\psi\rangle\otimes|0\rangle)=L^2|\psi\rangle\otimes|-2\rangle+(LR+RL)|\psi\rangle\otimes|0\rangle+R^2|\psi\rangle\otimes|2\rangle$
and the probabilities of reaching sites $|-2\rangle$, $|0\rangle$  and $|2\rangle$ in two steps are given by $\Vert L^2|\psi\rangle\Vert^2$, $\Vert (LR+RL)|\psi\rangle\Vert^2$ and $\Vert R^2\vert\psi\rangle\Vert^2$, respectively. The calculations work in a similar way for larger times.

\medskip

More recently, the model of {\bf Open Quantum Random Walks} (OQW) has been proposed by Attal et al. \cite{attal}. Let $\{B_{ij}\}_{i,j=1,\dots,k}$ belong to $M_d(\mathbb{C})$, the order $d$ square complex matrices, such that for each $j=1,\dots,k$,
\begin{equation}
\sum_{i=1}^k B_{ij}^*B_{ij}=I,
\end{equation}
where $B^*$ denotes the adjoint of $B$ and $I$ denotes the order $d$ identity matrix. We say that $k$ is the number of {\bf sites} and $d$ is the {\bf degree of freedom} on each site. Define
\begin{equation}\label{dens_attal}
\rho:=\sum_{i=1}^k\rho_i\otimes |i\rangle\langle i|,\;\;\;\rho_i\in M_d(\mathbb{C}), \;\;\;\rho_i\geq 0, \;\;\;\sum_{i=1}^k Tr(\rho_i)=1,
\end{equation}
where $\rho_i\geq 0$ means that $\rho_i$ is positive semidefinite. For a given initial density matrix  of such form, the OQW on $k$ sites induced by the $B_{ij}$, $i,j=1,\dots,k$ is, by definition \cite{attal}, the quantum channel
\begin{equation}\label{oqrw_def}
\Phi(\rho):=\sum_{i=1}^k\Big(\sum_{j=1}^k B_{ij}\rho_j B_{ij}^{*}\Big)\otimes |i\rangle\langle i|.
\end{equation}
We say that $B_{ij}$ is the effect matrix of transition from site $j$ to site $i$.
The case of homogeneous nearest neighbor OQWs on $\mathbb{Z}$ can be written, for matrices satisfying $L^*L+R^*R=I$, in the form
\begin{equation}\label{oqrwbasexp}
\Phi(\rho)=\sum_{i\in\mathbb{Z}} (R\rho_{i-1}R^*+L\rho_{i+1}L^*)\otimes|i\rangle\langle i|.
\end{equation}
In a similar way as in closed walks, we have a statistical interpretation of the traces of the matrices associated to each site. As an example, for the nearest neighbor walk above, if $\rho^{(0)}=\rho\otimes |0\rangle\langle 0|$, then $\Phi(\rho^{(0)})=L\rho L^*\otimes|-1\rangle\langle -1|+R\rho R^*\otimes|1\rangle\langle 1|$ and the probability of reaching sites $|-1\rangle$ or $|1\rangle$ after one step are respectively $Tr(L\rho L^*)$ and $Tr(R\rho R^*)$. In two steps, we get
\begin{equation}
\Phi^2(\rho^{(0)})=L^2\rho L^{2*}\otimes|-2\rangle\langle -2|+(LR\rho R^*L^*+RL\rho L^*R^*)\otimes|0\rangle\langle 0|+R^2\rho R^{2*}\otimes|2\rangle\langle 2|,
\end{equation}
and the probabilities of reaching sites $|-2\rangle$, $|0\rangle$  and $|2\rangle$ in two steps, are given by $Tr(L^2\rho L^{2*})$, $Tr(LR\rho R^*L^*+RL\rho_0 L^*R^*)$  and $Tr(R^2\rho R^{2*})$, respectively. The calculations work in a similar way for larger times. This is one of the simplest examples of OQWs.

\medskip

It is worth noting that if we have a density of the form (\ref{dens_attal}) then $\Phi(\rho)=\sum \eta_i\otimes|i\rangle\langle i|$ for some $\eta_i\geq 0$ \cite{attal}. In particular, the projections $|i\rangle\langle i|$ do not get mixed, so the channel describing the walk, when restricted to densities of such form, can be seen as acting on a direct sum space.

\medskip

Since the publication of \cite{attal}, several works have appeared describing some of the statistical properties of OQWs. We mention some of the topics studied so far:
\begin{enumerate}
\item Central Limit Theorems for OQWs \cite{attal2,sadowski}.
\item Reducibility, periodicity, ergodic properties of OQWs \cite{carboneaihp}.
\item Large Deviations for OQWs \cite{carbonejstatp}.
\item Microscopic derivation of open quantum Brownian motion \cite{sinayskiy4}.
\item Criteria for site recurrence of OQWs on $\mathbb{Z}$ \cite{ls2015}.
\item Hitting times for OQWs \cite{ls2016}.
\end{enumerate}

\medskip

See \cite{cfmhtf} for a more complete list of publications. In this work we are interested in further studying item 5 above, namely, recurrence properties of OQWs and we proceed with this goal having in mind some related properties of UQWs. One of the notions of recurrence for iterative unitary dynamics that we consider here has been defined by Gr\"unbaum et al. \cite{bourg,werner}. The other one, which is also discussed here in the context of OQWs, has been presented by \v Stefa\v n\'ak et al. \cite{stefanak}. The notion of monitored recurrence for OQWs studied in this work has been first described in \cite{ls2015} and is closely motivated by the unitary notion as in \cite{bourg,werner}.

\medskip

We remark that the dynamics of OQWs are in general quite different from the coined UQWs. The main reason for this is simple and is highlighted by the following observations:

\medskip
\begin{enumerate}
\item In the context of OQWs, we are interested in {\bf iterative quantum trajectories}. By this we mean that we prepare an initial density matrix, apply the channel on it and then perform a measurement to determine the site for which the system has evolved to. Then by normalizing we obtain a new density and we repeat the process. So, at each step we have a probability computation and this procedure is thus identified with {\bf summing the squares of amplitudes}. This typically produces gaussian curves or linear combinations of these as probability distributions \cite{attal}. See Figure 1, left.
\item In the context of coined unitary quantum random walks, it is well-known that the probability distribution is a very particular one and this happens due to interference: we let the system evolve and at the $n$-th step we perform a measurement. In other words, we {\bf sum amplitudes and then take the square modulus}. The typical outcome is the Konno distribution \cite{kempe}. See Figure 1, right.
\end{enumerate}

\begin{figure}[ht]
\begin{center}
\includegraphics[scale=0.5]{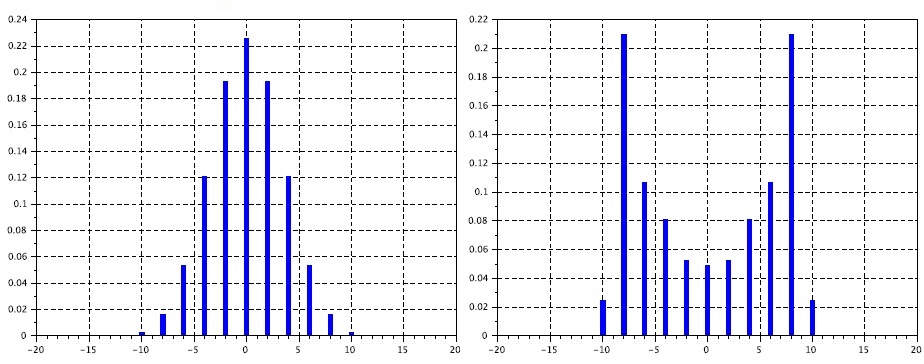}
\caption{\footnotesize{
Probability of visiting a site at time $n=12$ for the OQW (left graph) and the associated UQW (right graph) induced by matrices (\ref{splithad}). If in the UQW we perform measurements for each possible path (instead of measuring the sum of the amplitudes at the $n$-th step only), then such walk produces the same distribution as the associated OQW. The distribution in a) approaches a Gaussian curve \cite{attal} and the one in b) approaches the Konno distribution \cite{kempe}.
In both cases the walk begins at site $|0\rangle$, with the balanced distribution $\frac{1}{\sqrt{2}}[ 1 \; i ]^T$ and probabilities at odd-numbered sites being equal to zero. If the initial distribution is not balanced, then the distribution obtained for the UQW may be asymmetric but still fundamentally different from the OQW case.}}
\end{center}
\end{figure}

In this work, and with these observations in mind, we address the following problems:

\subsection{Recurrence criteria for OQWs}

What are the conditions on $L$ and $R$ so that the homogeneous nearest neighbor OQW on $\mathbb{Z}$ is site recurrent? A notion of recurrence for OQWs has been defined in \cite{ls2015}:

\medskip

{\bf Definition.} Let $\Phi$ be an OQW. We say that a {\bf site} $|i\rangle$ is {\bf recurrent} with respect to $\Phi$ if for every density matrix $\rho_i$ the following holds: beginning at $\rho_i\otimes|i\rangle\langle i|$, the probability of ever returning to site $|i\rangle$ equals 1. In the notation of hitting times, let $\pi_r(i;j)$ be the set of all products of $r$ matrices corresponding to a sequence of vertices that a walk is allowed to perform with $\Phi$, starting at $|i\rangle$ and first reaching $|j\rangle$ in the $r$-th step. Note that $\pi_r(i,j)\cap \pi_s(i,j)=\emptyset$ if $r\neq s$ (see [\cite{ls2016}, Section 5] for examples). Then site $|i\rangle$ is recurrent for $\Phi$ if, for every density matrix $\rho_i$,
\begin{equation}
\mathcal{R}_{|i\rangle}(\rho_i):=\sum_{r=1}^\infty\sum_{C\in\pi_r(i,i)} Tr(C\rho_i C^*)=1.
\end{equation}
We say that $\Phi$ is {\bf site recurrent} if every site is recurrent. This is a natural notion of site recurrence of quantum trajectories produced by an OQW \cite{ls2015}. We will sometimes refer to this notion as {\bf monitored} recurrence (see below the corresponding notion for unitary maps).

\medskip

{\bf Remark.} We emphasize that in the notions of recurrence studied in this work we are only interested in the return to the initial site, regardless of the associated density obtained at the moment of return (see Figure 2). Then we check if this return probability equals 1 for every given $\rho$. This is not the only possible definition, and we refer the reader to \cite{werner,ls2015,stefanak} for more on this matter.

\medskip

In \cite{ls2015}, a criterion for recurrence of OQWs is proved for certain walks on $\mathbb{Z}$. We restate such result:

\begin{theorem}\cite{ls2015}\label{teo2014}
Let $L$, $R$ be order 2 PQ-matrices inducing a homogeneous nearest neighbor unital OQW $\Phi$ on $\mathbb{Z}$. Then, the associated walk is recurrent if and only if the eigenvalues of $L^*L$ and $R^*R$ are equal to $1/2$.
\end{theorem}

We recall that in dimension 2, the PQ-matrices are just the diagonal or antidiagonal matrices \cite{ls2015}. In the original work, the theorem was stated in terms of the square moduli of the nonzero entries of the PQ matrices, but in dimension 2, these are equal to the eigenvalues of $L^*L$ and $R^*R$.
In this work we prove:
\begin{theorem}\label{rectheo}
Consider  a nearest neighbor OQW on $\mathbb{Z}$ induced by matrices $L$ and $R$. If the eigenvalues of $L^*L$ and $R^*R$ are equal to $1/2$, then the walk is recurrent. Moreover, the converse holds if the matrices are normal.
\end{theorem}

It is a simple matter to show that the converse of this theorem without the normality assumption may be false, namely, the OQW induced by the rows of the Hadamard matrix,
\begin{equation}\label{splithad}
R=\frac{1}{\sqrt{2}}\begin{bmatrix} 1 & 1 \\ 0 & 0 \end{bmatrix},\;\;\;L=\frac{1}{\sqrt{2}}\begin{bmatrix} 0 & 0 \\ 1 & -1 \end{bmatrix},
\end{equation}
produces a recurrent walk whose eigenvalues are equal to $0$ and $1$ (see Example \ref{bas_oqw_hadrec}). It is clear that there are cases in Theorem \ref{teo2014} which are not covered by Theorem \ref{rectheo} and conversely. Also in Section \ref{secanotherex}, we make a separate discussion for an example which does not meet the hypothesis of the above results.

\begin{figure}[ht]
\begin{center}
\includegraphics[scale=0.6]{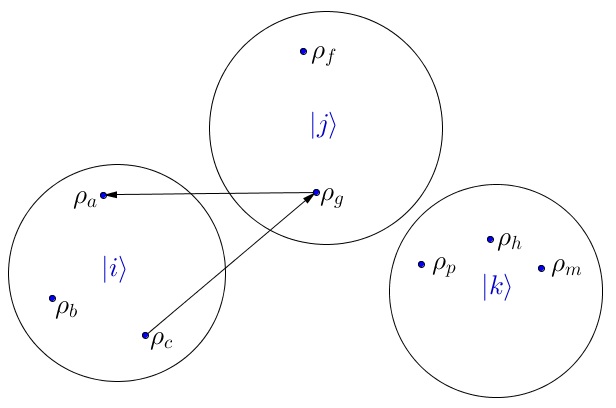}
\caption{\footnotesize{The circles correspond to sites and the $\rho_i$ inside correspond to densities on such site. We consider a notion of {\bf site} recurrence, that is, if we begin in a circle, will we ever return to it with probability 1? The arrows above illustrate a first return to site $|i\rangle$ in 2 steps. }}
\end{center}
\end{figure}

\medskip

We also remark that in the converse of the above theorem, the assumption that both $L$ and $R$ are normal implies unitality and the proof also holds if we assume unitality and that one of the matrices is normal. The proof of Theorem \ref{rectheo} is presented in Section \ref{secproofmain}.

\medskip

Regarding the OQW setting we recall that we have a Markov chain $(\rho_n,X_n)_{n\in\mathbb{N}}$ with values in $\mathcal{E}(\mathcal{H})\times I_k$, where $\mathcal{E}(\mathcal{H})$ is the set of density matrices on $\mathcal{H}$ Hilbert space, and $I_k=\{1,\dots,k\}$ accounts for the set of vertices of the graph. Then, from any position $(\rho,i)$, one can jump to one of the other values given by
\begin{equation}
\Big(\frac{1}{p(i,j)}B_{ji}\rho B_{ji}^{*},j\Big),
\end{equation}
with probability
\begin{equation}
p(i,j)=Tr(B_{ji}\rho B_{ji}^{*}).
\end{equation}
Then one might ask why we should look for recurrence criteria when a well established description already exists in the theory of Markov chains \cite{grimmett}. The answer to this is that besides being interested in recurrence criteria which are based essentially on the entries of the matrices, we remark that if we are given a pair $(\rho,X)$ (density in a given position), then as previously discussed, we are less interested in {\it state} recurrence than {\it site} recurrence. This corresponds to calculate the hitting probability of a collection of states. State recurrence still makes sense, but it is not the focus of this work.

\medskip

Another point that should be emphasized, which is discussed in \cite{stefanak}, is the fact that recurrence may depend on the initial state. Then we have a trichotomy:
\begin{enumerate}
\item A site of an OQW may be recurrent for every initial density matrix. Example: Hadamard OQW, and   this is proved in Example \ref{bas_oqw_hadrec}.
\item A site of an OQW may be recurrent for some, but not all densities. Example: A diagonal OQW such as the one induced by $L=diag(1/\sqrt{2},1/\sqrt{3})$, $R=diag(1/\sqrt{2},\sqrt{2}/\sqrt{3})$. To see this, consider the densities given by the matrix units $E_{11}=diag(1,0)$ (results in recurrence) and $E_{22}=diag(0,1)$ (transience).
\item A site of an OQW may be transient, i.e., it is not recurrent for any density. Example: once again a suitable diagonal choice suffices. Another example is the OQW induced by the amplitude damping channel, $p\in(0,1)$ \cite{ls2015}.
\end{enumerate}

\subsection{Recurrence for UQWs, P\'olya numbers and their application for OQWs}\label{ssaccessi}

We recall that in Gr\"unbaum et al. \cite{bourg,werner}, the authors discuss notions of state and site recurrence of iterative unitary evolutions in terms of {\bf monitoring}. Mathematically, {\it monitoring} a system is different from {\it measuring} a system: if $U$ is a unitary map and $|\psi\rangle$ is a initial state, we may inspect the recurrence of $|\psi\rangle$ in the $n$-th step. This is made by iterating $U$ once and then projecting onto the complement of the space generated by $|\psi\rangle$ (the monitoring procedure). We repeat this procedure and at the $n$-th step we iterate one more step, project onto $|\psi\rangle$ and then perform a measurement (i.e., take the square modulus of the amplitude function). Following \cite{bourg}, write
\begin{equation}a_{n,U,V}=a_n=PU\tilde{U}^{n-1}P,\;\;\;\tilde{U}_V=\tilde{U}=(I-P)U,\end{equation}
where $P=P_V$ is the orthogonal projection onto some subspace $V$ of $\mathcal{H}$. For $\psi\in V$, we say that
\begin{equation}R(|\psi\rangle)=\sum_{n\geq 1}\Vert a_n|\psi\rangle\Vert^2\end{equation}
is the {\bf $V$-return probability} of $\psi$.

\medskip

{\bf Definition.} Let $U$ be a unitary map on a Hilbert space $\mathcal{H}$, $V$ a subspace of $\mathcal{H}$ and $|\psi\rangle\in V$ a unit vector. We say that the state $|\psi\rangle$ is {\bf $V$-recurrent} if $R(|\psi\rangle)=1$; that is, $|\psi\rangle$ is recaptured by $V$ with probability 1. We say that a subspace $V$ is {\bf recurrent} if every state $|\psi\rangle$ in $V$ is $V$-recurrent. As in the OQW case, this is also called {\bf monitored} state/subspace recurrence.

\medskip

Then we see that {\bf site recurrence} can also be described in this notation: given a state $|\psi\rangle\otimes|x\rangle$ on site $|x\rangle$, we would like to know whether it is recaptured by (the space generated by) site $|x\rangle$ with probability 1, that is, whether it is $V_x$-recurrent, where
\begin{equation}V_x=span\{|\uparrow\rangle\otimes|x\rangle,|\downarrow\rangle\otimes|x\rangle\},\;\;\;x\in\mathbb{Z}.\end{equation}

\medskip

We also recall that in \cite{stefanak} a different notion of recurrence of unitary systems is presented. In this work, we review such notion (also revisited in \cite{werner}) and later we show that it can also be used in open systems. Let $p_0(n;\psi)=p_0(n)$ denote the probability of reaching site $|0\rangle$ given that the walk started at $|\psi\rangle\otimes |0\rangle$ and let
\begin{equation}\ov{P}_k=\prod_{n=1}^k[1-p_0(n)]\end{equation}
denote the probability of not finding the particle at the origin in the first $k$ trials. We define
\begin{equation}\ov{P}=\prod_{n=1}^\infty[1-p_0(n)]\end{equation}
in an analogous way. Then ${P}_k=1-\ov{P_k}$ is the probability of finding the particle at least once at the origin in the first $k$ trials. The number ${P}=1-\ov{P}$ is defined accordingly, and it is called the {\bf P\'olya number} of the UQW. We have the following equivalence:
\begin{equation}S:=\sum_{n=0}^\infty p_0(n)=\infty\;\Longleftrightarrow\; \prod_{n=1}^\infty[1-p_0(n)]=0\end{equation}
(that is, $S$ diverges if and only if $\ov{P}$ equals zero).

\medskip

{\bf Definition.}\cite{stefanak} We say that a UQW walk is {\bf SJK-recurrent} with respect to state $\psi$ if and only if $P=P(\psi)=1$.

\medskip

The term above is named after the authors \cite{stefanak}. In \cite{werner} it is shown an example where, for iterative unitary evolutions, SJK-recurrence is not equivalent to monitored recurrence. How about this notion for OQWs? In Section \ref{secpolyapatheq}, noting that an analogous definition of P\'olya numbers for OQWs is given, we prove the following:

\begin{theorem}\label{polyagback}
For an OQW, we have the following: a) If for each $\rho$, $(\rho,i)$ is monitored-recurrent then for each $\rho$, $(\rho,i)$ is SJK-recurrent. b) If a site is SJK-recurrent for a given OQW with respect to some state $(\rho,i)$ then $|i\rangle$ is monitored-recurrent with respect to some state $(\rho',i)$ which is accessible from $(\rho,i)$.
\end{theorem}
Above, to say that a state $\rho'$ is {\bf accessible} from $\rho$ means that there is a path on the graph, described by a product of transition matrices of the OQW, e.g. $B=B_{i_{n-1}}^iB_{i_{n-2}}^{i_{n-1}}\cdots B_{i_1}^{i_2}$, such that $\rho'=B\rho B^*/Tr(B\rho B^*)$. We also note that even though the converse b) above is apparently weaker than the direct implication, we will see that in many cases this is enough to prove monitored-recurrence of the walk (see the end of Section \ref{secpolyapatheq} and Section \ref{secanotherex}).

\subsection{An interference term: relating recurrence of UQW and OQWs}\label{lemarecsec}

First we note that any two matrices $L$, $R$ of the form
\begin{equation}\label{gooddecomp}
R=\begin{bmatrix} a & b \\0 & 0 \end{bmatrix},\;\;\; L=\begin{bmatrix} 0 & 0 \\ c & d \end{bmatrix}\end{equation}
such that a) $U=L+R$ is unitary and b) $L^*L+R^*R=I$, induce both a unitary and an open quantum walk on $\mathbb{Z}$ given by (\ref{canonicuqw}) and (\ref{oqrwbasexp}), respectively. Whenever both a UQW and an OQW are considered, we assume in this work that this is the decomposition being used. One basic example is given by walk induced by the rows of the Hadamard matrix \ref{splithad} and we call the associated walks the {\bf unitary} (resp. open) {\bf Hadamard walk}.
In Section \ref{secinterf} we show that the information of recurrence of the OQW and UQW induced by a pair of matrices is connected by an interference term:
\begin{theorem}\label{i_theoo2}
If $R_u(\psi)$ and $R_o(\psi)$ denote, respectively, the monitored recurrence probability for the UQW and OQW induced by a pair of matrices, both with respect to a given initial state $|\psi\rangle$, then
\begin{equation}
R_u(\psi)=R_o(\psi)+\alpha(\psi)
\end{equation}
where $\alpha(\psi)$ is an interference term.
\end{theorem}
The following result is a simple consequence:
\begin{cor}\label{interf_coro}
Let $L$, $R$ be two matrices satisfying $L^*L+R^*R=I$ and such that $U=L+R$ is unitary. Suppose that $\alpha(\psi)\leq 0$ (the so-called negative interference assumption). Let $\Phi$ denote the nearest neighbor homogeneous OQW on $\mathbb{Z}$ induced by $L$ and $R$, and let $U_\Phi$ denote the UQW associated to $\Phi$. Let $V=V_{|x\rangle}$ be the vector space associated to site $|x\rangle$. If $|\psi\rangle\otimes |x\rangle$ is V-recurrent for the UQW $U_\Phi$, then the pure density $\rho_{\psi,x}=|\psi\rangle\langle\psi|\otimes |x\rangle\langle x|$ is V-recurrent for the OQW $\Phi$. As a consequence, if site $|x\rangle$ is recurrent for the UQW, then it is recurrent for the OQW with respect to any pure density $\rho_{\psi,x}$.
\end{cor}

Following \cite{werner}, we can also present a simple relation between OQWs on the line and the Schur function of the associated QRW (see Example \ref{plancherelex}).

\medskip

At this point at least one important remark is in order, namely, the negative interference assumption $\alpha(\psi)\leq 0$ is certainly not always valid. Moreover, the problem of being able to control or estimate such function is a nontrivial one. Can one obtain natural, concrete examples, for which such assumption is true? We show in this work that the unitary coined Hadamard walk on $\mathbb{Z}$ in fact satisfies such assumption, thus producing an interesting relation between the open and closed walks induced by the Hadamard matrix.

\medskip

In the Appendix, as a complement to a result on positive recurrence studied in \cite{ls2015}, we prove the following.

\begin{theorem}\label{oqkactheo}(Open quantum Kac's lemma). Consider an irreducible recurrent OQW and assume that $|x\rangle$ is a positive recurrent site with associated density $\rho_x$. Let $E_R(\rho_x)$ denote the expected return time of the OQW with initial density $\rho_x$. If there is a unique stationary measure $\pi=\sum_i \pi(i)\otimes|i\rangle\langle i|$ associated to $\rho_x$, then
\begin{equation}
\rho_{st,\rho_x}(j)=E_R(\rho_x)\pi(j)
\end{equation}
and
\begin{equation}
E_R(\rho_x)=\frac{1}{Tr(\pi(x))}.
\end{equation}
\end{theorem}
The relevant definitions and the simple proof of the theorem are given in the Appendix. We remark that Theorem \ref{oqkactheo} is not applicable for homogeneous OQWs on $\mathbb{Z}$, due to the absence of the mentioned stationary measure in such case. We refer the reader to \cite{sinkovicz2} for a related result.

\medskip

{\bf Remark.} With the exception of the Appendix, we note that all examples considered in this work concern homogeneous nearest neighbor walks on $\mathbb{Z}$. However, many of the facts presented still hold if we consider general graphs, or different coins on different sites, with analogous proofs. We also refer the reader to \cite{xiong} for a discussion on the decomposition of order $n$ unitary matrices into a sum of row matrices and its properties (the expression (\ref{gooddecomp}) being the case $n=2$).

\section{Preliminaries: quantum channels and representation matrices}

OQWs are particular examples of completely positive maps and we refer the reader to \cite{bhatia,petz} for basic definitions. We recall that completely positive maps can be written in the Kraus form \cite{petz}, namely, there exist certain matrices $B_i$ such that
\begin{equation}\label{krausform}
\Phi(\rho)=\sum_i B_i\rho B_i^*.
\end{equation}
We say that $\Phi$ is {\bf trace-preserving} if $Tr(\Phi(\rho))=Tr(\rho)$ for all $\rho\in M_d(\mathbb{C})$, a condition which is equivalent to $\sum_i B_i^*B_i=I$. We say that $\Phi$ is {\bf unital} if $\Phi(I)=I$, which is equivalent to $\sum_i B_iB_i^*=I$. Trace-preserving completely positive (CPT) maps are also called {\bf quantum channels}. Also recall that if $A\in M_d(\mathbb{C})$ there is the corresponding vector representation $vec(A)$ associated to it, given by stacking together the matrix rows. For instance, if $d=2$,
\begin{equation}
A=\begin{bmatrix} a_{11} & a_{12} \\ a_{21} & a_{22}\end{bmatrix}\;\;\;\Rightarrow \;\;\; vec(A)=\begin{bmatrix} a_{11} & a_{12} & a_{21} & a_{22}\end{bmatrix}^T.
\end{equation}
The $vec$ mapping satisfies $vec(AXB^T)=(A\otimes B)vec(X)$ for any $A, B, X$ square matrices \cite{hj2}, with $\otimes$ denoting the Kronecker product. In particular, $vec(BXB^*)=vec(BX\ov{B}^T)=(B\otimes \ov{B})vec(X)$,
from which we can obtain the {\bf matrix representation} $[\Phi]$ for the CP map (\ref{krausform}):
\begin{equation}\label{matrep}
[\Phi]=\sum_{i} B_{i}\otimes \ov{B_{i}}=\sum_{i,j,k,l}\langle E_{kl},\Phi(E_{ij})\rangle E_{ki}\otimes E_{lj}
\end{equation}
We refer the reader to \cite{ls2015} for examples of matrix representation of well-known quantum channels. It should be clear that the spectrum of the channel is given by the corresponding information extracted from $[\Phi]$ and this will be useful later in this work. We recall the well-known fact that the matrix representation of a CPT map $\Phi: M_d(\mathbb{C})\to M_d(\mathbb{C})$ is independent of the Kraus representation considered. The proof of this result is a simple consequence of the unitary equivalence of Kraus matrices for a given quantum channel \cite{petz}.

\section{Relating recurrence for UQWs and OQWs}\label{secinterf}

First note that for any vectors $|a\rangle$ and $|b\rangle$,
\begin{equation}
\Vert|a\rangle\otimes|b\rangle\Vert^2=(|a\rangle\otimes|b\rangle)^*(|a\rangle\otimes|b\rangle)=\langle a|a\rangle\langle b|b\rangle=\Vert |a\rangle\Vert^2\Vert |b\rangle\Vert^2,
\end{equation}
(see \cite{hj2} ex. 28, p. 253), which implies
\begin{equation}\label{eqauxxx}
\Vert C|\psi\rangle\otimes|0\rangle\Vert^2=\Vert C|\psi\rangle\Vert^2
\end{equation}
and
\begin{equation}\label{tracenormequiv}
\Vert C|\psi\rangle\otimes|0\rangle\Vert^2=\langle\psi|C^*C|\psi\rangle\langle 0|0\rangle=\langle\psi|C^*C|\psi\rangle=Tr(C|\psi\rangle\langle\psi|C^*).
\end{equation}
Suppose that we wish to determine the recurrence of a UQW walk on $\mathbb{Z}$ with initial state $|\psi\rangle\otimes|0\rangle$. Recall from the Introduction that $\pi_r(i,i)$ is the collection of all paths allowed by the matrices of the OQW, starting at $|i\rangle$ and first returning to $|i\rangle$ in the $r$-th step. Then we have
$$R(|\psi\rangle\otimes|0\rangle)=\sum_{n\geq 1}\Vert a_{2n}|\psi\rangle
\Vert^2=\sum_{n\geq 1}\Big[\sum_{C\in\pi_n(0,0)} \Vert C|\psi\rangle\Vert^2+2\sum_{{C,D\in\pi_n(0,0), C\neq D}}Re\langle C|\psi\rangle, D|\psi\rangle\rangle\Big]$$
\begin{equation}=\sum_{n\geq 1}\sum_{C\in\pi_n(0,0)} \Vert C|\psi\rangle\Vert^2+\sum_{n\geq 1}\alpha_n(\psi),\end{equation}
where
\begin{equation}
\alpha_n(\psi)=2\sum_{{C,D\in\pi_n(0,0), C\neq D}}Re\langle C|\psi\rangle, D|\psi\rangle\rangle.\end{equation}
Hence,
\begin{equation}\label{oqwappears}
R(|\psi\rangle\otimes|0\rangle)=\sum_{n\geq 1}\Vert a_{2n}|\psi\rangle
\Vert^2=\sum_{n\geq 1}\sum_{C\in\pi_n(0,0)} \Vert C|\psi\rangle\Vert^2+\sum_{n\geq 1}\alpha_n(\psi).
\end{equation}
Note that by (\ref{eqauxxx}) and (\ref{tracenormequiv}), the double summation term in (\ref{oqwappears}) is exactly the calculation of first return probability of an OQW. Then we conclude: the recurrence probability of the unitary walk equals the recurrence probability of the open walk plus an interference term. This interference may lead to recurrence of the unitary walk which is larger or smaller than its open counterpart.
If we let $R_u(\psi)$ denote the recurrence probability for the UQW and if
\begin{equation}R_o(\psi)=\sum_{n\geq 1}\sum_{C\in\pi_n(0,0)} \Vert C|\psi\rangle\Vert^2\end{equation}
denotes the recurrence probability for the OQW, then by defining $\alpha(\psi):=\sum_{n\geq 1}\alpha_n(\psi)$, we get
\begin{equation}
R_u(\psi)=R_o(\psi)+\alpha(\psi).
\end{equation}
As a conclusion, if we understand the behavior of $\alpha_n$ then we are able to conclude certain facts:
\begin{enumerate}
\item First, note that $0\leq R_u,R_o\leq 1$. Therefore, if an OQW is recurrent we have $R_o=1$ and so by the above equation we must have $\alpha(\psi)\leq 0$. Moreover, if $\alpha(\psi)=0$, then the UQW is also recurrent, else it must be transient.
\item Consider the particular case with a negative interference,
\begin{equation}\label{special_cond}
\alpha_n(\psi)\leq 0,\;\;\;n=1,2,\dots
\end{equation}
Then,
\begin{equation}R(|\psi\rangle\otimes|0\rangle)\leq \sum_{n\geq 1}\sum_{C\in\pi_n(0,0)} \Vert C\psi\Vert^2=R_o(\psi),\end{equation}
as the expression on the RHS of this inequality corresponds to the probability of return of quantum trajectories beginning at site $|0\rangle$ with initial density $\rho=|\psi\rangle\langle\psi|$. Therefore, the validity of inequality (\ref{special_cond}) allows us to conclude: recurrence of the monitored unitary system starting at $|\psi\rangle\otimes |0\rangle$ implies recurrence of the quantum trajectory of the associated open walk starting at the pure state $|\psi\rangle\langle\psi|\otimes|0\rangle\langle 0|$. Conversely, under the assumption (\ref{special_cond}), if we know that a certain OQW is not recurrent with respect a certain pure density then we also know that the corresponding UQW is not recurrent for any state corresponding to that density. We register these facts in Corollary \ref{interf_coro}, presented in the Introduction.
\end{enumerate}

\begin{example}\label{plancherelex}{\bf Plancherel theorem and OQWs.} In \cite{werner}, the authors describe a relation between Schur functions and the problem of recurrence of iterative unitary evolutions on Hilbert space. The Schur function $f$ is also closely related to the Stieltjes moment generating function and we can write
\begin{equation}
f(z)=\frac{1}{z}\frac{\ov{\widehat{\mu}}(z)-1}{\ov{\widehat{\mu}}(z)},\;\;\;\widehat{\mu}(z)=\sum_{n=0}^\infty \mu_nz^n,\;\;\;\mu_n=\langle \psi|U^n\psi\rangle
\end{equation}
Following the mentioned authors, fix $r<1$, consider the series $\hat{a}(re^{it})=\sum_n r^na_ne^{itn}$ as a Fourier series and then obtain from the Plancherel theorem,
\begin{equation}\sum_n r^{2n}|a_n|^2=\frac{r^2}{2\pi}\int_{-\pi}^\pi|f(re^{it})|^2dt\end{equation}
where $a_n$ represents the amplitude associated to a monitored first return to the origin at step $n$. Taking the limit $r\to 1$, we obtain
\begin{equation}R_u(\psi)=\sum_n|a_n|^2=\frac{1}{2\pi}\lim_{r\to 1}\int_{-\pi}^\pi|f(re^{it})|^2dt,\end{equation}
and by the identity $R_u=R_o+\alpha$, we have
\begin{equation}R_o(\psi)=\frac{1}{2\pi}\lim_{r\to 1}\int_{-\pi}^\pi|f(re^{it})|^2dt-\alpha(\psi),\end{equation}
so we see that recurrence of the OQW is dependent on the interference pattern produced by the UQW and its associated Schur function $f$. So, in principle, the fact that $f$ is inner may not be enough to ensure (or forbid) recurrence of the associated OQW. On the other hand, by [\cite{werner}, Theorem 1], if $\alpha(\psi)<0$ then the spectral measure $\mu_\psi$ for the UQW has an absolutely continuous part and so the unitary walk is not recurrent with respect to $|\psi\rangle$. Once again, it is essential to have some knowledge on the behavior of the interference pattern.
\end{example}
\qee

\section{Hadamard walk on the integers: open and unitary versions}\label{secinterf2}

Here we illustrate some of the differences between the UQW and the OQW on $\mathbb{Z}$, both induced by the Hadamard coin via (\ref{canonicuqw}) and (\ref{oqrwbasexp}), respectively.

\begin{example}
(Probabilities for the UQW and OQW Hadamard walks on $\mathbb{Z}$). Let
\begin{equation}\label{hada_example}
R=\frac{1}{\sqrt{2}}\begin{bmatrix} 1 & 1 \\ 0 & 0 \end{bmatrix}, \;\;\;L=\frac{1}{\sqrt{2}}\begin{bmatrix} 0 & 0 \\ 1 & -1 \end{bmatrix}, \;\;\;C=L+R=\frac{1}{\sqrt{2}}\begin{bmatrix} 1 & 1 \\ 1 & -1 \end{bmatrix},\;\;\;|\psi\rangle=|\downarrow\rangle\otimes|0\rangle=\begin{bmatrix}0 \\ 1 \end{bmatrix}\otimes|0\rangle.
\end{equation}
Then, one step of the associated UQW is given by
\begin{equation}U|\psi\rangle=R|\psi\rangle\otimes|1\rangle+L|\psi\rangle\otimes|-1\rangle=\frac{1}{\sqrt{2}}\begin{bmatrix} 1 \\ 0 \end{bmatrix}\otimes|1\rangle-\frac{1}{\sqrt{2}}\begin{bmatrix} 0 \\ 1 \end{bmatrix}\otimes|-1\rangle.\end{equation}
Let $X_n$ denote the position of the particle, under the action of the UQW, at time $n$ with the initial $|\psi\rangle$ fixed above. Then, by routine calculations, we can obtain probabilities of reaching a particular site, given an initial state. For instance,
$$U^3|\psi\rangle=L^3|\psi\rangle\otimes|-3\rangle+(L^2R+RL^2+LRL)|\psi\rangle\otimes|-1\rangle
+(LR^2+R^2L+RLR)|\psi\rangle\otimes|1\rangle+R^3|\psi\rangle\otimes|3\rangle$$
\begin{equation}=\begin{bmatrix} 0 \\ -\frac{\sqrt{2}}{4}\end{bmatrix}\otimes |-3\rangle+
\begin{bmatrix} \frac{\sqrt{2}}{4} \\ -\frac{\sqrt{2}}{2} \end{bmatrix}\otimes|-1\rangle+\begin{bmatrix}  0 \\ \frac{\sqrt{2}}{4} \end{bmatrix}\otimes|1\rangle
+\begin{bmatrix} \frac{\sqrt{2}}{4} \\ 0\end{bmatrix}\otimes |3\rangle,\end{equation}
which implies the probabilities (with the initial $|\psi\rangle$ above)
\begin{equation}P(X_3=-3)=\frac{1}{8},\;\;\;P(X_3=-1)=\frac{5}{8},\;\;\;P(X_3=1)=\frac{1}{8},\;\;\;P(X_3=3)=\frac{1}{8}. \end{equation}
In the case of OQWs, we have to consider quantum trajectories and, as such, we have to perform  measurements at each step so the calculations produce probabilities which are different from the unitary case. In other words, instead of taking the square modulus of a sum of amplitudes, we are summing probabilities, as discussed in the Introduction. If $Y_n$ denotes the position of the particle under the action of the OQW at time $n$ with the initial $|\psi\rangle$ fixed above then, for $\rho=|\psi\rangle\langle\psi|$,
\begin{equation}P(Y_3=-3)=Tr(L^3\rho L^{3*})=\frac{1}{8},\;\;\;P(Y_3=3)=Tr(R^3\rho R^{3*})=\frac{1}{8},\end{equation}
\begin{equation}P(Y_3=-1)=Tr(LLR\rho R^*L^*L^*)+Tr(RLL\rho L^*L^*R^*)+Tr(LRL\rho L^*R^*L^*)=\frac{3}{8},\end{equation}
\begin{equation}P(Y_3=1)=Tr(RRL\rho L^*R^*R^*)+Tr(LRR\rho R^*R^*L^*)+Tr(RLR\rho R^*L^*R^*)=\frac{3}{8}.\end{equation}
As expected, the probabilities for the OQW and UQW on time $n=3$ are distinct, and for larger times the UQW produces the well-known Konno distribution, whereas the OQW produces a gaussian distribution.
\end{example}

\qee

Now we note that in the case of the Hadamard walk, an induction argument shows that for every $|\psi\rangle$,
\begin{equation}C|\psi\rangle=\begin{bmatrix}\pm 1/2^{2n} \\ 0 \end{bmatrix} \textrm{ or } \begin{bmatrix} 0 \\ \pm 1/2^{2n} \end{bmatrix},\;\;\;C\in\pi_{2n}(0,0),\end{equation}
the signal and position of the nonzero entry depending on the particular sequence. Then
\begin{equation}R(\psi)=\sum_{n\geq 1}\Big[\sum_{C\in\pi_{2n}(0,0)} \Vert C|\psi\rangle\Vert^2+2\sum_{{C,D\in\pi_{2n}(0,0), C\neq D}}Re\langle C\psi, D\psi\rangle\Big]<\sum_{n\geq 1}\sum_{C\in\pi_{2n}(0,0)} \Vert C|\psi\rangle\Vert^2,\end{equation}
that is, we have nonpositive interference for every initial state $|\psi\rangle$. We omit the proof. Below, we show a table with the first steps of monitored unitary and open dynamics, comparing the square modulus of a sum of amplitudes (probability corresponding to UQWs) and the sum of square modulus (probability corresponding to OQWs). As an example we see that, for any initial density, the probability that a unitary monitored walk starting at site $|i\rangle$ first returns to $|i\rangle$ in $6$ steps equals zero, a fact which does not hold if we let the system evolve without monitoring.

\begin {table}[ht]
\begin{center}
\begin{tabular}{|c|c|c|}
\hline
$\# \textrm{ Steps}$  & $\textrm{OQW: }\sum \Vert a_{2k}\psi\Vert^2$ & $\textrm{UQW: } \Vert\sum a_{2k}\psi\Vert^2$  \\
\hline
$2$ & $1/2$ & $1/2$ \\
\hline
$4$ & $1/2^3$ & $1/2^3$ \\
\hline
$6$ & $1/2^4$ & $0$ \\
\hline
$8$ & $5/2^7$ & $1/2^7$ \\
\hline
$10$ & $7/2^8$ & $0$ \\
\hline
$12$ & $21/2^{10}$ & $2/2^{10}$ \\
\hline
$14$ & $33/2^{11}$ & $0$ \\
\hline
$16$ & $429/2^{15}$ & $25/2^{15}$ \\

\hline

\end{tabular}
\caption {Probability of first return to $0$ at time $n=2k$ for the Hadamard OQW and UQW. Note that such probabilities do not depend on the initial state $\psi$ at site $|0\rangle$.}
\end{center}
\end{table}

Therefore, we are able to conclude that the UQW, with any initial state $|\psi\rangle$, is not recurrent. On the other hand, in the following example we give a direct proof that the associated OQW is site recurrent.

\begin{example}\label{bas_oqw_hadrec}(Site recurrence of the Hadamard OQW on $\mathbb{Z}$).
Consider once again the matrices $L$ and $R$ from (\ref{hada_example}).
It is easy to show that the nearest neighbor OQW on $\mathbb{Z}$ induced by these matrices is recurrent. In fact, let us consider the case $i=0$, the other cases being identical. First recall that the number of paths of length $2k$ starting at $0$ and first returning to $0$ at time $n=2k$ is \cite{ls2015}
\begin{equation}\label{path_count_exp}
\alpha_{2k}=\frac{1}{2k-1}{2k\choose k}.
\end{equation}
Second, it is a simple matter to see that for all $n\geq 1$, if $\rho=(r_{ij})$ is a density matrix, then
\begin{equation}Tr(L^{l_n}R^{r_n}\cdots L^{l_1}R^{r_1}\rho R^{r_1*}L^{l_1*}\cdots R^{r_n*}L^{l_n*})=\frac{1}{2^{\sum_i (l_i+r_i)}}(1-2Re(r_{12})),\;\;\; r_1>0,\end{equation}
that is, for any sequence of $L$'s and $R$'s, we have a specific expression for the trace whenever the first conjugation of the density $\rho$ is $R\rho R^*$ (i.e. considering composition of conjugations we have that $R\rho R^*$ occurs first). We also check that for all $n\geq 1$,
\begin{equation}Tr(R^{r_n}L^{l_n}\cdots R^{r_1}L^{l_1}\rho L^{l_1*}R^{r_1*}\cdots L^{l_n*}R^{r_n*})=\frac{1}{2^{\sum_i (l_i+r_i)}}(1+2Re(r_{12})),\;\;\; l_1>0,\end{equation}
that is, we obtain a slightly different expression if the first conjugation is with matrix $L$. Noting that there are $\alpha_{2k}/2$ ways of leaving and returning  to $0$ if we are always to the right (or left) of $0$ and using (\ref{path_count_exp}) we see that the probability of ever returning to $0$ is given by
\begin{equation}R=\sum_{k=1}^\infty \Big( \frac{\alpha_{2k}}{2}\frac{1}{2^{2k}}(1+2Re(r_{12})+1-2Re(r_{12}))\Big)=\sum_{k=1}^\infty\alpha_{2k}\frac{1}{2^{2k}}=\sum_{k=1}^\infty\frac{1}{2k-1}{2k\choose k}\frac{1}{4^k}=1.\end{equation}
This proves that starting at any site $|i\rangle$, we certainly return to it (with probability 1), regardless of the initial density matrix $\rho\otimes|i\rangle\langle i|$.

\end{example}

\qee

\section{SJK-recurrence and Fourier analysis for OQWs}

In this section we study the Fourier transform of a nearest neighbor OQW on $\mathbb{Z}$. This will allow us to examine SJK-recurrence in the open setting and we present examples in this section and in Section \ref{secanotherex}. We note that our analysis is in part related to the ones presented in \cite{konno,stefanak}.
Let $K=(-\pi,\pi]$ and $\mathcal{F}:\bigoplus_{j\in\mathbb{Z}} M_2\to L^2(K,M_2)$ be given by
\begin{equation}
(\mathcal{F}\rho)(k)=\widehat{\rho}(k):=\sum_{j\in\mathbb{Z}} e^{-ijk}\rho(j),\;\;\;k\in K,\;\;\;\rho=\sum_{j\in\mathbb{Z}} \rho(j)\otimes|j\rangle\langle j|\in\bigoplus_{j\in\mathbb{Z}} M_2.
\end{equation}
We are interested in $\mathcal{F}$ acting on density matrices. We have that $\mathcal{F}^{-1}:L^2(K,M_2)\to\bigoplus_{j\in\mathbb{Z}} M_2$ is thus given by
\begin{equation}
\mathcal{F}^{-1}(f)=\sum_{j\in\mathbb{Z}}\frac{1}{2\pi}\int_Ke^{ijk}f(k)\;dk\otimes|j\rangle\langle j|,\;\;\;f\in L^2(K,M_2).
\end{equation}
Given an OQW $\Phi$, we are interested in the map $\Lambda$ such that the following diagram commutes:
\begin{equation}\label{cdd}
\begin{CD}\bigoplus_{j\in\mathbb{Z}} @>\Phi>> \bigoplus_{j\in\mathbb{Z}}  \\
@V{\mathcal{F}}VV  @V{\mathcal{F}}VV \\
L^2(K,M_2) @>{\Lambda}>> L^2(K,M_2)
\end{CD}
\end{equation}
We call $\Lambda=\Lambda_{\Phi}$ the Fourier transform of the OQW $\Phi$. By definition, if $f\in L^2(K,M_2)$, then
\begin{equation}\Lambda(f)=\mathcal{F}\circ\Phi\circ\mathcal{F}^{-1}(f)=\mathcal{F}\circ\Phi\circ\Big(\sum_{j\in\mathbb{Z}}\frac{1}{2\pi}\int_Ke^{ijk}f(k)\;dk\otimes|j\rangle\langle j| \Big)=\mathcal{F}\circ\Phi\circ\Big(\sum_{j\in\mathbb{Z}}\eta_{\rho}(j)\otimes|j\rangle\langle j| \Big),\end{equation}
where
\begin{equation}\eta_{\rho}(j)=\frac{1}{2\pi}\int_Ke^{ijk}f(k)\;dk.\end{equation}
Then,
\begin{equation}\mathcal{F}\circ\Phi\circ\Big(\sum_{j\in\mathbb{Z}}\eta_{\rho}(j)\otimes|j\rangle\langle j| \Big)=\mathcal{F}\circ\Big(\sum_{j\in\mathbb{Z}}[R\eta_{\rho}(j-1)R^*+L\eta_{\rho}(j+1)L^*]\otimes|j\rangle\langle j| \Big).\end{equation}
Therefore,
\begin{equation}\Lambda(f)(k)=\frac{1}{2\pi}\sum_{j\in\mathbb{Z}} e^{-ijk}\Big[L\Big(\int_Ke^{i(j+1)l}f(l)\;dl\Big)L^*+R\Big(\int_Ke^{i(j-1)l}f(l)\;dl\Big)R^*\Big],\;\;\;f\in L^2(K,M_2)\end{equation}

\begin{example}

Suppose that $f(k)=F$ for all $k$, where $F\in M_2$ is a fixed matrix. Then, since
\begin{equation}\int_Ke^{i0k}dk=\int_K\;dk=2\pi,\;\;\;\int_Ke^{ijk}dk=0,\;\;\;j\neq 0,\end{equation}
we get
\begin{equation}\sum_{j\in\mathbb{Z}} e^{-ijk}\int_Ke^{i(j+1)l}\;dl=2\pi e^{ik}\end{equation}
and
\begin{equation}\sum_{j\in\mathbb{Z}} e^{-ijk}\int_Ke^{i(j-1)l}\;dl=2\pi e^{-ik}.\end{equation}
So,
$$\Lambda(f)(k)=\frac{1}{2\pi}\Bigg( L\Big[\sum_{j\in\mathbb{Z}} e^{-ijk}\int_Ke^{i(j+1)l}F\;dl\Big]L^*+R\Big[\sum_{j\in\mathbb{Z}} e^{-ijk}\int_Ke^{i(j-1)l}F\;dl\Big]R^*\Bigg)$$
\begin{equation}=\frac{1}{2\pi}\Big(2\pi e^{ik}LFL^*+2\pi e^{-ik}RFR^*\Big)=e^{ik}LFL^*+e^{-ik}RFR^*.\end{equation}
A similar calculation gives
\begin{equation}\Lambda^2(f)(k)=e^{2ik}L^2FL^{*2}+(RLFL^*R^*+LRFR^*L^*)+e^{-2ik}R^2FR^{*2}.\end{equation}
Above we see that the sites are indexed by some complex exponential $e^{-ijk}$ associated to site $j$ and such expression codifies all the ways that one can move in two steps, starting from site 0.
\end{example}

\qee

Now, define
\begin{equation}\rho^{(n)}:=\sum_x \rho_x^{(n)}\otimes |x\rangle\langle x|\;\Longrightarrow\; \mathcal{F}(\rho^{(n)})(k)=\widehat{\rho^{(n)}}(k)=\sum_x e^{-ikx}\rho_x^{(n)},\end{equation}
a density at time $n$ and its Fourier transform. Note that we can write
\begin{equation}\widehat{\rho^{(n+1)}}(k)=\widehat{\Phi}(k)[\widehat{\rho^{(n)}}(k)]\end{equation}
where
\begin{equation}\widehat{\Phi}(k)(F)=e^{ik}LFL^*+e^{-ik}RFR^*,\;\;\;F\in M_2(\mathbb{C}).\end{equation}
If we set
\begin{equation}L=\begin{bmatrix} a & b \\ c & d\end{bmatrix},\;\;\;R=\begin{bmatrix} x & y \\ z & w\end{bmatrix},\end{equation}
then we are interested in the eigenvalues and eigenvectors of the matrix representation of $\widehat{\Phi}(k)$, which is
\begin{equation}[\widehat{\Phi}(k)]=e^{ik}[L]+e^{-ik}[R]=\begin{bmatrix}
e^{ik}|a|^2+e^{-ik}|x|^2 & e^{ik}a\ov{b}+e^{-ik}x\ov{y} &
e^{ik}\ov{a}b+e^{-ik}\ov{x}y & e^{ik}|b|^2+e^{-ik}|y|^2 \\
e^{ik}a\ov{c}+e^{-ik}x\ov{z} & e^{ik}a\ov{d}+e^{-ik}x\ov{w} & e^{ik}b\ov{c}+e^{-ik}y\ov{z} & e^{ik}b\ov{d}+e^{-ik}y\ov{w} \\
e^{ik}c\ov{a}+e^{-ik}z\ov{x} & e^{ik}c\ov{b}+e^{-ik}z\ov{y} & e^{ik}d\ov{a}+e^{-ik}w\ov{x} & e^{ik}d\ov{b}+e^{-ik}w\ov{y} \\
e^{ik}|c|^2+e^{-ik}|z|^2 & e^{ik}c\ov{d}+e^{-ik}z\ov{w} &
e^{ik}\ov{c}d+e^{-ik}\ov{z}w & e^{ik}|d|^2+e^{-ik}|w|^2
\end{bmatrix}.\end{equation}
Denote by $\lambda_j(k)$ and $v_j(k)$, $j=1,2,3,4$, the eigenvalues and corresponding eigenvectors of $\Phi(k)$. Under the assumption that the eigenvectors produce an orthonormal eigenbasis for the space of matrices (e.g. the channel is normal), we are able to write
$$\widehat{\rho^{(n)}}(k)=\lambda_1(k)^n\langle v_1(k),\widehat{\rho^{(0)}}(k)\rangle v_1(k)+\lambda_2(k)^n\langle v_2(k),\widehat{\rho^{(0)}}(k)\rangle v_2(k)$$
\begin{equation}+\lambda_3(k)^n\langle v_3(k),\widehat{\rho^{(0)}}(k)\rangle v_3(k)+\lambda_4(k)^n\langle v_4(k),\widehat{\rho^{(0)}}(k)\rangle v_4(k),\end{equation}
that is,
\begin{equation}\label{rhofourier}
\widehat{\rho^{(n)}}(k)=\sum_{j=1}^4\lambda_j(k)^n\langle v_j(k),\widehat{\rho^{(0)}}(k)\rangle v_j(k).
\end{equation}
Now we can perform the inverse Fourier transform and obtain the exact expression for the probability amplitudes. For each site we have
\begin{equation}\label{rhotimen}
\rho_x^{(n)}=\frac{1}{2\pi}\int_K e^{ixk}\widehat{\rho^{(n)}}(k)dk.
\end{equation}
In analogy to the classical problem, we restrict to walks which start at the origin. Therefore, the initial condition is
\begin{equation}\rho^{(0)}=\rho_0\otimes |0\rangle\langle 0|.\end{equation}
Its Fourier transform is then
\begin{equation}\mathcal{F}(\rho^{(0)})(k)=\widehat{\rho^{(0)}}(k)=\sum_x e^{-ikx}\rho_x^{(0)}=\rho_0.\end{equation}
Then, (\ref{rhofourier}) becomes
\begin{equation}
\widehat{\rho^{(n)}}(k)=\sum_{j=1}^4\lambda_j(k)^n f_j(k)=\sum_{j=1}^4\lambda_j(k)^n\langle \rho_0,v_j(k)\rangle v_j(k).
\end{equation}
Therefore, from (\ref{rhotimen}), the probability of return to $0$ at time $n$ is
\begin{equation}\label{probformula1}
p_0(n)=Tr(\rho_0^{(n)})=Tr\Big(\frac{1}{2\pi}\int_K\widehat{\rho^{(n)}}(k)dk\Big).
\end{equation}

\medskip

{\bf Remark.} We note that if we follow Konno \cite{konno}, we can write
\begin{equation}\label{konnodualeq}
p_x^{(n)}=\frac{1}{2\pi}\int_K e^{ixk}Tr(\rho_0 Y_n(k))dk,\;\;\;Y_n(k)=(e^{ik}L_{B^*}R_B+e^{-ik}L_{C^*}R_C)^n(I),
\end{equation}
where $L_B(X):=BX$, $R_B(X):=XB$, $B,X\in M_n(\mathbb{C})$ and so
\begin{equation}p_0^{(n)}=\frac{1}{2\pi}\int_K Tr(\rho_0 Y_n(k))dk.\end{equation}

\medskip

Now, combining expressions (\ref{rhofourier})
and (\ref{probformula1}), we get
\begin{equation}\label{probexp1}
p_0(n)=\sum_{j=1}^4\int_K\lambda_j(k)^n\langle v_j(k),\widehat{\rho^{(0)}}(k)\rangle Tr(v_j(k))\frac{dk}{2\pi}=\sum_{j=1}^4\int_K\lambda_j(k)^n Tr( v_j(k)^*\widehat{\rho^{(0)}}(k))Tr(v_j(k))\frac{dk}{2\pi}=\sum_{j=1}^4 I_j(n),
\end{equation}
where
\begin{equation}\label{probexp2}
I_j(n)=\int_K\lambda_j(k)^n f_j(k)\frac{dk}{2\pi},\;\;\;f_j(k)=Tr( v_j(k)^*\widehat{\rho^{(0)}}(k))Tr(v_j(k)).
\end{equation}

\begin{example}\label{bfgenex}
(Bit-flip OQW). Consider the nearest neighbor bit-flip OQW on $\mathbb{Z}$ \cite{ls2015} induced by the Kraus matrices
\begin{equation}
L=\sqrt{p}\begin{bmatrix} 1 & 0 \\ 0 & 1 \end{bmatrix},\;\;\;R=\sqrt{1-p}\begin{bmatrix} 0 & 1 \\ 1 & 0 \end{bmatrix},\;\;\;p\in(0,1).
\end{equation}
By the above calculations, we can write
\begin{equation}
[\widehat{\Phi}(k)]=e^{ik}[L]+e^{-ik}[R]=\begin{bmatrix}pe^{ik} & 0 & 0 & (1-p)e^{-ik} \\ 0 & pe^{ik} & (1-p)e^{-ik} & 0 \\ 0 & (1-p)e^{-ik} & pe^{ik} & 0 \\ (1-p)e^{-ik} & 0 & 0 & pe^{ik}\end{bmatrix}.
\end{equation}
We have that the eigenvalues for $\widehat{\Phi}$ are
\begin{equation}\lambda_1(k)=\lambda_2(k)=pe^{ik}+(1-p)e^{-ik},\;\;\;\lambda_3(k)=\lambda_4(k)=pe^{ik}-(1-p)e^{-ik},\end{equation}
with corresponding eigenvectors given by
\begin{equation}
v_1(k)=\frac{1}{\sqrt{2}}\begin{bmatrix} 1 \\ 0 \\0 \\ 1\end{bmatrix},\;\;\;
  v_2(k)=\frac{1}{\sqrt{2}}\begin{bmatrix} 0 \\ 1 \\1 \\ 0\end{bmatrix},\;\;\;
  v_3(k)=\frac{1}{\sqrt{2}}\begin{bmatrix} 1 \\ 0 \\0 \\ -1\end{bmatrix},\;\;\;
  v_4(k)=\frac{1}{\sqrt{2}}\begin{bmatrix} 0 \\ 1 \\-1 \\ 0\end{bmatrix}.
\end{equation}
Note that $Tr(vec^{-1}v_1)=\sqrt{2}$ and $Tr(vec^{-1}v_i)=0$, $i=2,3,4$.
Then (\ref{probexp2}) simplifies to just one term and (\ref{probexp1}) becomes
\begin{equation}p_0(n)=\int_K\lambda_1(k)^n Tr( v_1(k)^*\widehat{\rho^{(0)}}(k))Tr(v_1(k))\frac{dk}{2\pi}=\int_K(pe^{ik}+(1-p)e^{-ik})^n Tr( v_1(k)^*\widehat{\rho^{(0)}}(k))\sqrt{2}\frac{dk}{2\pi}.\end{equation}
Now, if $\widehat{\rho}^{(0)}(k)=\rho_0$, then $Tr( v_1(k)^*\widehat{\rho^{(0)}}(k))=1/\sqrt{2}Tr(\rho_0)=1/\sqrt{2}$ and so the above becomes
\begin{equation}p_0(n)=\int_K(pe^{ik}+(1-p)e^{-ik})^n\frac{dk}{2\pi}=\left\{
\begin{array}{ll}
\binom{n}{\frac{n}{2}}p^{n/2}(1-p)^{n/2} & \textrm{ if } n \textrm{ is even.}  \\
0 & \textrm{ if } n \textrm{ is odd.}
\end{array} \right.
\end{equation}
A simple calculation with the binomial theorem (also see [\cite{grad}, eq. 0.241(3)]) shows that
\begin{equation}
\sum_{n=1}^\infty p_0(n)=\frac{1}{|1-2p|}-1<\infty,\;\;\;p\neq\frac{1}{2}.
\end{equation}
Therefore, for $p\neq 1/2$ we conclude that such bit-flip OQW is SJK-transient. As for the case $p=1/2$, note that
\begin{equation}p_0(n)=\int_K\frac{1}{2^n}(e^{ik}+e^{-ik})^n\frac{dk}{2\pi}=\int_K\frac{1}{2^n}(2\cos(k))^n\frac{dk}{2\pi}=\int_K\cos^n(k)\frac{dk}{2\pi}.\end{equation}
Such integral equals zero if $n$ is odd, and in the even case, we note that for $n=0,1,2,\dots$,
\begin{equation}\label{solvingcosine}
p_0(2n)=\int_K\cos^{2n}(k)\frac{dk}{2\pi}=\binom{2n}{n}\frac{1}{4^n},
\end{equation}
and since
\begin{equation}\label{cosineresposta}
\sum_{n}{2n\choose n}\frac{1}{4^n}=\infty,
\end{equation}
we conclude that the nearest neighbor OQW on the line induced by the bit-flip channel is SJK-recurrent if, and only if, $p=1/2$. Note that this is the same criterion for monitored recurrence of the bit-flip OQW obtained by [\cite{ls2015}, Theorem 4.6], which can be applied here, since the bit-flip is a PQ-channel.
\end{example}
\qee

\section{On the equivalence of monitored recurrence and SJK-recurrence for OQWs}\label{secpolyapatheq}

Given an OQW, we are interested in studying the probability of reaching a given site, regardless of the associated density. We note that the OQW dynamics of the position alone is not a Markov chain, since it must also be determined by the density matrix degree of freedom. Nevertheless, the reasoning followed in this section is a variation of known results coming from the classical theory of Markov chains \cite{grimmett}, but taking in consideration an extra degree of freedom at each site (i.e., the density matrix). Given a quantum trajectory $\omega=\{\omega_i=(\rho_i(\omega),X_i(\omega))\}_{i\in\mathbb{N}}\in (D(\mathcal{H})\times\mathbb{Z})^\mathbb{N}$, the time of first visit to state $i$ is
\begin{equation}
T_{i}(\omega)=\inf\{n\geq 1:X_i(\omega)=i\}
\end{equation}
The r-th passage time $T_{i}^{(r)}$ to state $i$ defined inductively as $T_{i}^{(0)}(\omega)=0$, $T_{i}^{(1)}(\omega)=T_{i}(\omega)$ and for $r=1,2,\dots$, we set
\begin{equation}
T_{i}^{(r+1)}(\omega)=\inf\{n\geq T_{i}^{(r)}(\omega)+1:X_n(\omega)=i\}
\end{equation}

{\bf Remark.} We note that the random variable $T_i:(D(\mathcal{H})\times\mathbb{Z})^\mathbb{N}\to\mathbb{N}\cup\{\infty\}$ is a stopping time for the quantum trajectory, since $\{T_i=n\}$ depends  only on $\omega_1=(\rho_1(\omega),X_1(\omega)), \omega_2=(\rho_2(\omega),X_2(\omega)),\dots, \omega_n=(\rho_n(\omega),X_n(\omega))$ (actually, it clearly depends only on the $X_1,\dots X_n$). In particular we have that $T_i$ is the first passage time of a set of elements from $(D(\mathcal{H})\times\mathbb{Z})^\mathbb{N}$ (and not just one element). This is an important difference between the quantum trajectory setting and a classical one.

\medskip

In the above language, to say that $|i\rangle$ is monitored-recurrent means that, for each $\rho$,
\begin{equation}\label{recassump}
P_{(\rho,i)}(T_i<\infty)=1.
\end{equation}
(the subscript $(\rho,i)$ above indicates the initial density and site). Since the above equation holds for every density, this implies that for each $\rho$ and for each $r$,
\begin{equation}\label{reccomplem1}
P_{(\rho,i)}(T_i^{(r)}<\infty)=1,
\end{equation}
since the $r$-th return behaves like the first return but with a different density.

\medskip

Denote by $V_i$ the number of visits to site $|i\rangle$. If $\omega\{\omega_n=(\rho_n,X_n)\}_{n\geq 0}$ denotes a quantum trajectory, we can write
\begin{equation}V_i(\omega)=\sum_{n=0}^\infty 1_{\{X_n=i\}}(\omega)\end{equation}
and note that
\begin{equation}\label{aax0}
\sum_{n=0}^\infty P_{(\rho,i)}(X_n=i)=\sum_{n=0}^\infty E_{(\rho,i)}(1_{\{X_n=i\}})=E_{(\rho,i)}\sum_{n=0}^\infty 1_{\{X_n=i\}}=E_{(\rho,i)}(V_i)
\end{equation}
Also note that
\begin{equation}\{\omega: \omega_0=(\rho,i) \textrm{ and } T_i^{(r)}<\infty\}=\{\omega: \omega_0=(\rho,i) \textrm{ and } V_i>r\}\end{equation}
In particular,
\begin{equation}\label{aax1}
P_{(\rho,i)}(V_i>r)=P_{(\rho,i)}(T_i^{(r)}<\infty)
\end{equation}
Then, under the monitored-recurrence assumption (\ref{recassump}) (which implies (\ref{reccomplem1})), and using (\ref{aax0}), (\ref{aax1}), we have that
\begin{equation}\sum_{n=0}^\infty P_{(\rho,i)}(X_n=i)\stackrel{(\ref{aax0})}{=}E_{(\rho,i)}(V_i)=\sum_{r=0}^\infty P_{(\rho,i)}(V_i>r)\stackrel{(\ref{aax1})}{=}\sum_{r=0}^\infty P_{(\rho,i)}(T_i^{(r)}<\infty)\stackrel{(\ref{reccomplem1})}{=}\infty,\;\;\;\forall\rho\end{equation}
where in the last equality we have used that (\ref{recassump}) and (\ref{reccomplem1}) holds for {\it every} $\rho$; note that it is not enough to assume that the return is certain for just one specific density. We have proved the following:
\begin{pro}\label{pro51} If, for each $\rho$, monitored recurrence holds for $(\rho,i)$ then, for each $\rho$, SJK-recurrence holds for $(\rho,i)$.\end{pro}
The need for the assumption of recurrence for {\it all} $\rho$ should be clear: one must have control on the recurrence properties of the site with respect to $\rho$ and all the densities which are accessible from it.

\medskip

Let $\mathcal{P}(\rho,i)$ denote the set of all densities in $i$ which are accessible starting from $(\rho,i)$ (recall the notion of accessibility given in the Introduction). Suppose $\sum_n p_i(n)=\infty$ for some $\rho$. Then by the classical Markov chain theory, it is immediate that the converse of Proposition \ref{pro51} holds for classical OQWs (transitions which are multiples of the identity). For the general case we proceed as follows. Let $\rho$ be any state on $i$ and suppose that no state in $\mathcal{P}(\rho,i)$ is monitored-recurrent. Let $f_i(\rho)=max_{\rho'\in\mathcal{P}(\rho,i)} P_{(\rho',i)}(T_i<\infty)<1$. Then, note
that
\begin{equation}P_{(\rho,i)}(V_i>r)=P_{(\rho,i)}(T_i^{(r)}<\infty)\leq f_i(\rho)^r,\end{equation}
and so
\begin{equation}\sum_{n=0}^\infty P_{(\rho,i)}(X_n=i)=E_{(\rho,i)}(V_i)=\sum_{n=0}^\infty P_{(\rho,i)}(V_i>r)\leq \sum_{n=0}^\infty f_i(\rho)^r=\frac{1}{1-f_i(\rho)}<\infty,\end{equation}
a contradiction. Hence, there must be a monitored-recurrent state in $\mathcal{P}_i(\rho)$. We summarize this result in the following:
\begin{pro}\label{lemmaaux1}
If a site $|i\rangle$ is SJK-recurrent for a given OQW with respect to some state $\rho$ then $|i\rangle$ is monitored-recurrent with respect to some state accessible from $\rho$.\end{pro}
The above proposition raises a natural question: if a site is SJK-recurrent for a given OQW with respect to some state $\rho$, then is it monitored-recurrent with respect to $\rho$ itself? The difficulty in answering the question lies on whether the OQW considered is recurrent with respect to all initial densities, or with respect to just some of them. Because of this, the answer is in principle negative in general. Under some additional conditions, we may be able to answer the above question positively; for instance, if it is known that the probability of first returns do not depend on the initial density. In this case, the OQW satisfies a dichotomy, that is, either all initial states produce a transient or a recurrent walk (see Section \ref{secanotherex} for one such example).


\section{Proof of Theorem \ref{rectheo}}\label{secproofmain}

In this section we prove the criterion for recurrence of homogeneous nearest neighbor OQWs on the integer line stated in the Introduction. We consider two matrices $L$ and $R$ such that $L^*L+R^*R=I$ which dictate the transition for all sites $\{|i\rangle:i\in\mathbb{Z}\}$ via expression (\ref{oqrwbasexp}). The analogous problem of obtaining a criterion for walks on $\mathbb{Z}^n$, $n>1$ is, up to our knowledge, unsolved. In the proof of the theorem we make use of the following inequalities.

\begin{lemma}\cite{wkh} Let $K, S$ be order $n$ matrices. If $K\geq 0$ and $S=S^*$, then
\begin{equation}
\lambda_{min}(S)Tr(K)\leq Tr(KS)\leq\lambda_{max}(S)Tr(K).
\end{equation}
\end{lemma}

\begin{lemma}\cite{marshall}
For arbitrary order $n$ matrices $X$ and $Y$, if $\sigma_{max}(\cdot)$ and $\sigma_{min}(\cdot)$ denote the largest and smallest singular value, respectively, then
\begin{equation}
\sigma_{min}(X)\sigma_{min}(Y)\leq\sigma_{min}(XY)\leq \sigma_{max}(XY)\leq\sigma_{max}(X)\sigma_{max}(Y).
\end{equation}
\end{lemma}

\medskip

{\bf Proof of Theorem \ref{rectheo}.} Suppose that one wishes to calculate the probability of recurrence of site $|0\rangle$ by a path counting argument. Recall that $\pi_{2r}(0,0)$ denotes the collection of all paths starting at $|0\rangle$ and first returning to it in $2r$ steps. By the invariance of this walk with respect to translations (i.e., we have transitions given by $L$ and $R$ on all sites), we may assume $|i\rangle=|0\rangle$. The probability that the walk ever returns to $|0\rangle$ is then
\begin{equation}
\mathcal{R}_{|0\rangle}(\rho)=\sum_{r=1}^\infty\sum_{C\in\pi_{2r}(0,0)} Tr(C\rho C^*)
\end{equation}
Now, recalling that $\alpha_{2r}=\frac{1}{2r-1}{2r\choose r}$ is the number of paths of length $2r$ starting at $|0\rangle$ and first returning to $|0\rangle$ at time $2r$, note that we always have the bounds
\begin{equation}\label{lowerbound1}
\sum_{r=1}^\infty\frac{1}{2r-1}{2r\choose r} \min_{C\in\pi_{2r}(0,0)} Tr(C\rho C^*)\leq \mathcal{R}_{|0\rangle}(\rho)\leq \sum_{r=1}^\infty\frac{1}{2r-1}{2r\choose r} \max_{C\in\pi_{2r}(0,0)} Tr(C\rho C^*)
\end{equation}
Now we show that we can provide certain estimates on the trace expression above in terms of the eigenvalues of $L$ and $R$. Note that, for every density matrix $\rho$, we have for $A=L^{l_1}R^{r_1}\cdots L^{l_m}R^{r_m}$, with $\sum_i l_i=\sum_i r_i=r$,
\begin{equation}
Tr(A\rho A^*)=Tr(\rho A^*A)\geq\lambda_{min}(A^*A)Tr(\rho)=\sigma_{min}(A)^2=\sigma_{min}(L^{l_1}R^{r_1}\cdots L^{l_m}R^{r_m})^2\geq\sigma_{min}(L)^{2r}\sigma_{min}(R)^{2r}.
\end{equation}
Analogously, $Tr(A\rho A^*)\leq\sigma_{max}(L)^{2r}\sigma_{max}(R)^{2r}$. Then, we obtain from (\ref{lowerbound1}) that
$$\sum_{r=1}^\infty\frac{1}{2r-1}{2r\choose r} \sigma_{min}(L)^{2r}\sigma_{min}(R)^{2r}\leq \sum_{k=1}^\infty\frac{1}{2r-1}{2r\choose r} \min_{C\in\pi_{2r}(0,0)} Tr(C\rho C^*)\leq \mathcal{R}_{|0\rangle}(\rho)$$
\begin{equation}\label{lowerbound2}
\leq\sum_{r=1}^\infty\frac{1}{2r-1}{2r\choose r} \max_{C\in\pi_{2r}(0,0)} Tr(C\rho C^*)\leq\sum_{r=1}^\infty\frac{1}{2r-1}{2r\choose r} \sigma_{max}(L)^{2r}\sigma_{max}(R)^{2r}
\end{equation}
Combine the assumption $\sigma_i(L)^2=\sigma_i(R)^2=\frac{1}{2}$ with the fact that $x\in(0,1) \mapsto\sum_{r=1}^\infty\frac{1}{2r-1}{2r\choose r}x^r(1-x)^r$ equals 1, its maximum, if and only if $x=1/2$. This proves the sufficiency. Now suppose that we have a recurrent nearest neighbor walk on $\mathbb{Z}$ and assume that $L, R$ are normal. First, we prove that for every $X$ we have
\begin{equation}\label{twosteps}
Tr(LRX R^*L^*)=Tr(RLX L^*R^*)
\end{equation}
In fact, note that the trace preservation assumption implies that $R^*R$ and $L^*L$ commute. For simplicity of notation, we assume that the matrices are of order 2, as the general case is proven with the same reasoning. Write $L^*L=UD_L U^*$ and $R^*R=UD_RU^*$, where $U$ is unitary, $D_L=diag(\lambda,\mu)$, $D_R=diag(1-\lambda,1-\mu)$, $0\leq \lambda,\mu\leq 1$. Write $U^*X U=\begin{bmatrix} x_{11} & x_{12} \\ x_{21} & x_{22}\end{bmatrix}$. Since $R$ is normal, we have
$$Tr(LRX R^*L^*)=Tr(L^*LRX R^*)=Tr([I-R^*R]RX R^*)=Tr(RX R^*)-Tr(R^*R^*RRX)$$
\begin{equation}
\stackrel{\textrm{R normal}}{=}Tr(R^*RX)-Tr(R^*RR^*RX)=Tr(D_RU^*X U)-Tr(D_R^2U^*X U)
\end{equation}
$$=(1-\lambda)x_{11}+(1-\mu)x_{22}-(1-\lambda)^2 x_{11}-(1-\mu)^2x_{22}=\lambda(1-\lambda)x_{11}+\mu(1-\mu)x_{22}$$
In a similar way as above,
\begin{equation}
Tr(RLX L^*R^*)=Tr(D_LU^*X U)-Tr(D_L^2U^*X U)=\lambda(1-\lambda)x_{11}+\mu(1-\mu)x_{22}
\end{equation}
This concludes the proof of (\ref{twosteps}), which is the calculation of first return in 2 steps. For 4 steps, we need to estimate, for instance, $Tr(LLRRX R^*R^*L^*L^*)$. Then, once again by the cyclic property of the trace and the normality assumptions, a simple calculation gives
$$Tr(LLRRX R^*R^*L^*L^*)=Tr(L^*L^*LLRRX R^*R^*)=Tr(L^*LL^*LRRX R^*R^*)=Tr((I-R^*R)(I-R^*R)RRX R^*R^*)$$
\begin{equation}
=Tr((I-R^*R)(I-R^*R)R^*RR^*RX)=Tr(D_R^2(I-D_R)^2U^*X U)=\lambda^2(1-\lambda)^2 x_{11}+\mu^2(1-\mu)^2(1-x_{11}).
\end{equation}
In general, under the normality assumptions, it is a simple matter to show that if $l_i, r_i\geq 0$ are integers, $\sum_i l_i=\sum_i r_i=r$, then
\begin{equation}\label{gen_exp_rec1}
Tr(L^{l_1}R^{r_1}\cdots L^{l_m}R^{r_m}X R^{r_m*}L^{l_m*}\cdots R^{r_1*}L^{l_1*})=\lambda^r(1-\lambda)^r x_{11}+\mu^r(1-\mu)^r(1-x_{11}),
\end{equation}
the above expression characterizing the probability of occurring a certain path of length $2k$ starting and ending at zero. In particular, for nearest neighbor walks on the integer line, the normality assumption implies that the probabilities for a path of length $k$ do not depend on the particular sequence. From (\ref{gen_exp_rec1}) we conclude that
\begin{equation}\label{final_exp_recnormal}
\mathcal{R}_{|0\rangle}(\rho)=\sum_{k=1}^\infty\sum_{C\in\pi_{2k}(0,0)} Tr(C\rho C^*)=\sum_{k=1}^\infty\frac{1}{2k-1}{2k\choose k}(\lambda^k(1-\lambda)^k x_{11}+\mu^k(1-\mu)^k(1-x_{11})),
\end{equation}
and recurrence of every pair $(|i\rangle,\rho)$ follows if and only if $\lambda=\mu=1/2$. To see this, note that if recurrence must hold for every density then we can reason on $\lambda$ and $\mu$ separately if we choose, with the $U$ above, a $\rho$ such that $U\rho U^*=E_{11}$ and $\rho'$ such that $U\rho'U^*=E_{22}$, respectively. This concludes the proof.

\qed

{\bf Remark.} If we suppose that both $L$ and $R$ are normal, then the unitality assumption follows. The proof also holds if we assume unitality and that one of the matrices is normal.

\section{Another example}\label{secanotherex}

The homogeneous OQW on $\mathbb{Z}$ induced by
\begin{equation}
L=\frac{1}{\sqrt{3}}\begin{bmatrix} 1 & 1 \\ 0 & 1 \end{bmatrix},\;\;\;R=\frac{1}{\sqrt{3}}\begin{bmatrix} 1 & 0 \\ -1 & 1 \end{bmatrix}
\end{equation}
has been studied previously in \cite{attal,carboneaihp,carbonejstatp,konno}. We have  $L^*L+R^*R=LL^*+RR^*=I$, but $LR\neq RL$ and $L$, $R$ are not normal. Also, it is easy to show that for this example $Tr(LRX R^*L^*)\neq Tr(RLX L^*R^*)$ in general, so the symmetry property (\ref{gen_exp_rec1}) appearing in the proof of Theorem \ref{rectheo} does not hold. Also note that $L$ and $R$ are not PQ-matrices \cite{ls2015}. A calculation gives that the eigenvalues of both $L^*L$ and $R^*R$ are
\begin{equation}\frac{1}{2}\pm\frac{\sqrt{5}}{6}.\end{equation}
A numerical experiment based on a path counting argument suggests that this walk is recurrent. Here we follow part of the analysis presented in \cite{konno} together with a calculation made in Example \ref{bfgenex} above in order to determine that the walk is in fact SJK-recurrent. First we write, for the OQW $\Phi$ induced by such matrices, the matrix representation of its Fourier transform,
\begin{equation}
[\widehat{\Phi}]=e^{ik}[L]+e^{-ik}[R]=\frac{1}{3}\begin{bmatrix}2\cos(k) & e^{ik} & e^{ik} & e^{ik} \\ -e^{-ik} & 2\cos(k) & 0 & e^{ik} \\ -e^{-ik} & 0 & 2\cos(k) & e^{ik} \\ e^{-ik} & -e^{-ik} & -e^{-ik} & 2\cos(k)\end{bmatrix},
\end{equation}
and we note that this corresponds to the transpose of the matrix representation of the dual process $Y_n(k)$ appearing in eq. (\ref{konnodualeq}). A calculation shows that the eigenvalues are
\begin{equation}\lambda_0=\frac{2}{3}\cos(k),\;\lambda_1=\frac{2}{3}\cos(k)+\frac{1}{3}\Big(\xi-\frac{1}{\xi}\Big),\end{equation}
\begin{equation}\lambda_2=\frac{2}{3}\cos(k)+\frac{1}{6}\Big[(-1+i\sqrt{3})\xi+\frac{1+i\sqrt{3}}{\xi}\Big],\;\lambda_3=\ov{\lambda_2},\end{equation}
where
\begin{equation}\xi=\xi(k)=(2\cos(k)+\sqrt{4\cos^2(k)+1})^{1/3},\end{equation}
in accordance with \cite{konno}. By writing $u=u(k)=\cos(k)$, we get
\begin{equation}\xi=\xi(u)=(2u+\sqrt{4u^2+1})^{1/3},\end{equation}
and if we let $s=s(u):=\xi(u)-1/\xi(u)$ a routine calculation gives that
\begin{equation}\lambda_1=\frac{1}{6}s(s^2+5).\end{equation}
\begin{figure}[ht]
\begin{center}
\includegraphics[scale=0.7]{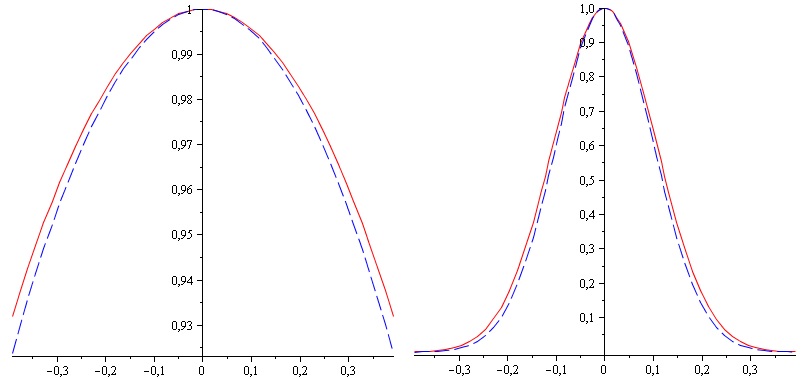}
\caption{\footnotesize{Left graph: $\cos(k)$ (dashed line) and $\lambda_1(k)$ (solid). Right graph: $\cos^{100}(k)$ (dashed line) and $\lambda_1^{100}(k)$. A cosine integral has appeared previously in Example \ref{bfgenex}, producing a general term for which the associated series is divergent.}}
\end{center}
\end{figure}
A simple analysis shows that $\lambda_1$ is the only eigenfunction reaching the extremes of the interval $[-1,1]$, and so the asymptotic behavior of $p_0^{(n)}$ is dictated by such eigenfunction. By [\cite{konno}, Theorem 4.10],
\begin{equation}\lim_{n\to\infty}\frac{p_0^{(2n)}}{\alpha_{2n}}=\frac{1}{\pi},\;\;\;\lim_{n\to\infty}\frac{p_0^{(2n+1)}}{\alpha_{2n+1}}=0,\end{equation}
where
\begin{equation}\alpha_{n}=\int_{-\pi/2}^{\pi/2}\lambda_1(k)^n dk.\end{equation}
So, the recurrence problem of the OQW is determined by the behavior of such integral. As noted by \cite{konno}, $\lambda_1(k)$ has a close resemblance to the cosine function, which has already appeared in Example \ref{bfgenex}. Therefore, the integral (\ref{solvingcosine}) can be compared with $\alpha_{2n}$ (see Figure 3) and the former has the explicit formula (\ref{cosineresposta}), which is the general term of a divergent series, so we conclude that the OQW considered in this example is SJK-recurrent with respect to every $\rho$. By Proposition \ref{lemmaaux1}, monitor recurrence holds with respect to some state accessible from $\rho$. To conclude that monitor recurrence holds for each densities, it is enough to show that the probability of ever returning to $i$, calculated via path counting, does not depend on any given density. We can show this by noting that if $X=X_1\cdots X_n$ is a matrix product of $L$'s and $R$'s for a path leaving and first returning to $i$ at time $n$, then so is $\sigma(X)=\sigma(X_1)\cdots\sigma(X_n)$, where $\sigma(L)=R$ and $\sigma(R)=L$. Then, it holds that $Tr(X\rho X^*)+Tr(\sigma(X)\rho \sigma(X)^*)$ does not depend on $\rho$. To see why, just use the fact that
\begin{equation}
L^{*n}L^n+R^{*n}R^n=\frac{n^2+2}{3^n}I,\;\;\;Tr(L^{*n}L^n)=Tr(R^{*n}R^n)=\frac{n^2+2}{3^n},
\end{equation}
(see \cite{konno}). Since we have previously shown that the walk is monitor recurrent with respect to some density, then it must be monitor recurrent with respect to every density, and we are done.

\section{Conclusions and further questions}

In this work we have discussed the problem of site recurrence of quantum walks on the integer line, giving emphasis to Open Quantum Random Walks, while comparing such model with the well-known unitary (coined) quantum random walk. Theorem \ref{rectheo} is one of the main results, describing recurrence of homogeneous OQWs in terms of spectral information of the transition matrices (the eigenvalues of $L^*L$ and $R^*R$). Also, in Theorem \ref{polyagback}, we are able to discuss a relation between monitored recurrence and recurrence via P\'olya numbers: if for each ``internal state'' $\rho$, $(\rho,i)$ is monitored-recurrent then for each $\rho$, $(\rho,i)$ is SJK-recurrent. Conversely, if a site is SJK-recurrent for a given OQW with respect to some state $(\rho,i)$ then $|i\rangle$ is monitored-recurrent with respect to some state $(\rho',i)$ which is accessible from $(\rho,i)$ (that is, there is a path on the graph given by a product of the matrices of the OQW, say~$B$, such that $\rho^\prime=B\rho B^*/Tr(B\rho B^*)$). 

\medskip

Theorem \ref{i_theoo2} concerns the quantities $R_u(\psi)$ and $R_o(\psi)$, the monitored recurrence probability for the UQW and OQW, respectively, induced by a pair of matrices $L$, $R$ (satisfying $L^*L+R^*R=I$ and such that $U=L+R$), both with respect to a given initial state $|\psi\rangle$. Then it is shown that $R_u(\psi)=R_o(\psi)+\alpha(\psi)$, $\alpha(\psi)$ the so-called interference term, and under certain assumptions on the interference, we are able to conclude recurrence of the OQW in terms of the recurrence of the associated UQW (Corollary \ref{interf_coro}), this topic being discussed in Sections \ref{secinterf} and \ref{secinterf2}. As a complement, Theorem \ref{oqkactheo} is an OQW version of Kac's theorem on the expected return time to a site, this result holding under the assumption of positive recurrence.

\medskip

An open question is the problem of obtaining site recurrence criteria for homogeneous OQWs on $\mathbb{Z}^d$, $d\geq 2$. Some of the ideas employed in this work may be of assistance in this generalization but, in principle, the problem seems to be more difficult and the case where the transition matrices are not normal may require a separate analysis. A related question is the problem of site recurrence of OQWs acting on other kinds of infinite graphs.

\medskip

In addition, and motivated in part by Example \ref{plancherelex}, a natural question concerns the study of site and subspace monitored-recurrence in terms of Carath\'eodory and Schur functions related to OQWs. More precisely, we would like to obtain the equivalent, for OQWs, to Theorem 2.4 in \cite{bourg}, which consists of a relation between the Schur function being inner and state/site recurrence of states in finite-dimensional subspaces. Note that this would hypothetically allow the determination of the interference term $\alpha(\cdot)$. We stress that this project seems to be far from trivial, since we still do not know how to define those functions (and if this is in fact possible) in this particular setting. 

\section{Appendix: positive recurrence revisited and Kac's theorem for OQWs}

In this appendix we review the notion of positive recurrence presented in \cite{ls2015} and prove Theorem \ref{oqkactheo}, a complement to the theory presented here.

\medskip

We say that a site $|y\rangle$ is {\bf accessible} from $|x\rangle$ if for every initial density at $|x\rangle$, there is a finite path going to $|y\rangle$ such that the trace of such path is strictly positive. Accessibility of a site $|y\rangle$ from $|x\rangle$ is denoted by $x\to y$ (compare the notion of {\it site} accessibility just given with the one of {\it state} accessibility defined in the end of Section \ref{ssaccessi}). We recall the following lemma.

\begin{lemma}\cite{ls2015} If $x$ is recurrent site and $x\to y$ then $y$ is recurrent with respect to every $\rho_y\otimes|y\rangle$ which is accessible by the given initial density $\rho_x\otimes|x\rangle$.
\end{lemma}

Now, let $\rho_x$ be any fixed density at site $x$ and write, for any given OQW with transition matrices $B_i^j$,
\begin{equation}S_{\rho_x,j}^1:=B_x^j\rho_x B_x^j\end{equation}
and
\begin{equation}S_{\rho_x,j}^n:=\sum_{i_1,\dots,i_{n-1}\neq x}B_{i_{n-1}}^j\cdots B_{i_1}^{i_2}B_x^{i_1}\rho_x B_x^{i_1*}B_{i_1}^{i_2*}\cdots B_{i_{n-1}}^{j*},\;\;\;n=2,3,\dots\end{equation}
This is the matrix consisting of the sum of all possible transitions from site $x$ (with initial density $\rho_x$) to site $j$ in $n$ steps and such that no visit to $x$ occurs during the first $n$ steps (except the case where $x=j$, when the first return occurs at the $n$-th step). If the degree of each vertex of the graph is finite, as it is usually assumed, this is a finite sum of products of $n$ matrices and is therefore well defined. By taking the trace, we obtain the probability of reaching site $j$ from $\rho_x\otimes|x\rangle$ in $n$ steps without intermediate visits to $x$. Now, define
\begin{equation}\rho_{st,\rho_x}(j):=\sum_{n=1}^\infty S_{\rho_x,j}^n.\end{equation}
The trace of such matrix corresponds to the expected time spent in $j$ between visits to $x$ (we do not count time $n=0$). In fact, if $T_x$ denotes the time of first return to $x$, that is,
\begin{equation}T_x=\inf\{n\geq 1:X_n=x\},\end{equation}
then
\begin{equation}Tr(\rho_{st,\rho_x}(j))=Tr(\sum_{n=1}^\infty S_{\rho_x,j}^n)=\sum_{n=1}^\infty P_{\rho_x}(X_n=j,T_x>n)=E_x\Big(\sum_{n=1}^{T_x-1}1_{X_n=j}\Big).\end{equation}
Finally, define
\begin{equation}\rho_{st,\rho_x}:=\sum_j\rho_{st,\rho_x}(j)\otimes |j\rangle\langle j|,\end{equation}
and the {\bf expected return time} by
\begin{equation}E_R(\rho_x):=Tr(\rho_{st,\rho_x})=\sum_j Tr(\rho_{st,\rho_x}(j)).\end{equation}

\medskip

{\bf Definition.}
We will say that $|x\rangle$ is a {\bf positive recurrent} site if it is recurrent and there exists $\rho_x$ such that \begin{equation}\label{tracofinito}
E_R(\rho_x)=\sum_j Tr(\rho_{st,\rho_x}(j))<\infty
\end{equation}
and
\begin{equation}\label{cond_imp}
\rho_{st,\rho_x}(x)=\sum_{n=1}^\infty S_{\rho_x,x}^n=\rho_x.
\end{equation}

We say that an OQW is {\bf irreducible} if for every $i,j$ distinct sites, we have $i\to j$ and $j\to i$ \cite{carboneaihp,ls2015}. Then we recall that  by [\cite{ls2015}, Theorem 5.8], if an OQW $\Phi$ on $\mathbb{Z}$ is recurrent and irreducible, then positive recurrence of the walk implies the existence of a stationary state for $\Phi$. However, we note that by [\cite{carbonejstatp}, Proposition 4.4], there are no invariant states for homogeneous OQWs on $\mathbb{Z}$. As a consequence, the open quantum version of Kac's theorem, as stated in the Introduction, is not applicable to such OQWs; one should instead consider, for instance, suitable nonhomogeneous walks on $\mathbb{Z}$ or finite graphs (both with a unique stationary measure).

\medskip

{\bf Proof of Theorem \ref{oqkactheo}.} By assumption we have that $E_R(\rho_x)<\infty$ for some $\rho_x$.  Now we note that $\rho_{st,\rho_x}$ is a stationary operator, by [\cite{ls2015}, Theorem 5.7]; therefore, by the assumed uniqueness, we have $\rho_{st,\rho_x}=c\pi$ for some $c$. By taking the trace of this equality and summing, we get
\begin{equation}E_R(\rho_x)=\sum_jTr(\rho_{st,\rho_x}(j))=c\sum_jTr(\pi(j))=c.\end{equation}

\medskip

Now, with the irreducibility and recurrence assumption, the fact that a stationary measure exists implies that the walk is positive recurrent. Therefore,
$\rho_{st,\rho_x}(x)=\rho_x$ and so $Tr(\rho_{st,\rho_x}(x))=Tr(\rho_x)=1$. Then
\begin{equation}\rho_{st,\rho_x}(j)=E_R(\rho_x)\pi(j)\;\Longrightarrow\; E_R(\rho_x)=\frac{1}{Tr(\pi(x))}.\end{equation}
\qed

\begin{example}{\bf OQW on $\mathbb{N}$ with retaining barrier.} We revisit an example seen in \cite{ls2015}. Let $B_0^1=I$ and for $i\geq 1$, let
\begin{equation}
B_{i}^{i-1}=\begin{bmatrix} \sqrt{q_{11}} & 0 \\ 0 & \sqrt{q_{22}}\end{bmatrix} ,\;\;\; B_i^{i+1}=\begin{bmatrix} \sqrt{p_{11}} & 0 \\ 0 & \sqrt{p_{22}}\end{bmatrix},
\end{equation}
with $p_{ii},\; q_{ii}\geq 0$, $p_{ii}<q_{ii}$, $i=1, 2$, and assume that
$$B_i^{i-1*}B_i^{i-1}+B_i^{i+1*}B_i^{i+1}=\begin{bmatrix} p_{11}+q_{11} & 0 \\ 0 & p_{22}+q_{22}\end{bmatrix}=I.$$
This example can be seen as two copies of a random walk on $\mathbb{N}$ with a retaining barrier at zero for which moving left has a larger probability than moving right. Fix an initial density $\rho=\rho_0\otimes|0\rangle\langle 0|$, $tr(\rho_0)=1$. In general, we have
$$B_{i_{n-1}}^{i_n}B_{i_{n-2}}^{i_{n-1}}\cdots B_{i_1}^{i_2}B_{0}^{i_1}\rho_0 B_{0}^{i_1*}B_{i_1}^{i_2*}\cdots B_{i_{n-2}}^{i_{n-1*}}B_{i_{n-1}}^{i_n*}=\begin{bmatrix}  p_{11}^{k-1}q_{11}^{n-k}\rho_{11} & \sqrt{p_{11}p_{22}}^{k-1}\sqrt{q_{11}q_{22}}^{n-k}\rho_{12} \\ \sqrt{q_{11}q_{22}}^{k-1}\sqrt{p_{11}p_{22}}^{n-k}\ov{\rho_{12}} & p_{22}^{k-1}q_{22}^{n-k}\rho_{22} \end{bmatrix},$$
where $k$ is the number of times the walk has moved right (note that above we write $p^{k-1}$, and not $p^k$, since the first move is to the right with probability one). We claim that such walk is positive recurrent. In fact, first note that we can pick for instance $\rho_x=E_{11}$, and by the above expression, it is such that $\rho_{st,\rho_x}(x)$ has only one nonzero entry, namely entry $(1,1)$. By a classical argument, this entry must be the probability of every returning to zero in a homogeneous walk on $\mathbb{N}$ where it is more likely to move left than right and this equals 1. Therefore, $\rho_{st,\rho_x}(x)=E_{11}=\rho_x$. Also $\sum_i \rho_{st,\rho_x}(i)<\infty$, since a left-biased classical walk with a barrier is known to be positive recurrent, see [\cite{grimmett}, Section 6.4]. We let $\alpha=p_{11}/q_{11}$, define $\pi_j=\alpha^j(1-\alpha)E_{11}$, so by Theorem \ref{oqkactheo},
\begin{equation}
E_R(\rho_x)=\frac{1}{Tr(\pi(x))}=\frac{1}{\alpha^x(1-\alpha)}.
\end{equation}
\end{example}
\qee

{\bf Acknowledgements.} The authors would like to thank Paolo Giulietti and Dagoberto Justo for discussions concerning topics of this work. S.L.C. has been partially supported by FAPEMIG (Universal project CEX-APQ-00554-13). C.F.L. is grateful for the hospitality of ICMC-USP S\~ao Carlos and for the financial support of the XX Brazilian School of Probability, during which part of this work was done.

\end{document}